\documentclass[twocolumn]{htl-author}

\usepackage[table]{xcolor}  

\usepackage{tabularx}
\usepackage{geometry}
\usepackage{adjustbox} 
\usepackage{caption}
\usepackage{array}
\usepackage{pgfplots}
\pgfplotsset{width=10cm,compat=1.15} 
\usepackage{pgfplotstable}
\usepackage{booktabs} 
\usepackage{colortbl} 

\definecolor{fadedgreen}{RGB}{217, 237, 216}
\definecolor{lightgreen}{RGB}{217, 237, 216}
\definecolor{mediumgreen}{RGB}{173, 214, 170}
\definecolor{darkgreen}{RGB}{123, 193, 120}
\definecolor{reviewresponse}{RGB}{0, 0, 0}
\definecolor{khaki}{rgb}{0.93, 0.93, 0.9}

\begin{document}

\title{iSurgARy: A mobile augmented reality solution for ventriculostomy in resource-limited settings}

\author{Zahra Asadi$^{1^*}$, Joshua Pardillo Castillo$^{1^*}$, Mehrdad Asadi$^{1}$, David S. Sinclair$^{2}$, Marta Kersten-Oertel$^{1}$}

\address{$^{1}$Gina Cody School of Engineering and Computer Science, Concordia University, Montreal, QC, Canada\\
$^{2}$Departments of Neurology and Neurosurgery, Montreal Neurological Institute \& Hospital, McGill University, Montreal, QC, Canada}

\abstract{Global disparities in neurosurgical care necessitate innovations addressing affordability and accuracy, particularly for critical procedures like ventriculostomy. This intervention, vital for managing life-threatening intracranial pressure increases, is associated with catheter misplacement rates exceeding 30\% when using a freehand technique. Such misplacements hold severe consequences including haemorrhage, infection, prolonged hospital stays, and even morbidity and mortality.  To address this issue, we present a novel, stand-alone mobile-based augmented reality system (iSurgARy) aimed at significantly improving ventriculostomy accuracy, particularly in resource-limited settings such as those in low- and middle-income countries. iSurgARy uses landmark based registration by taking advantage of Light Detection and Ranging (LiDaR) to allow for accurate surgical guidance. To evaluate iSurgARy, we conducted a two-phase user study. Initially, we assessed usability and learnability with novice participants using the System Usability Scale (SUS), incorporating their feedback to refine the application. In the second phase, we engaged human-computer interaction (HCI) and clinical domain experts to evaluate our application, measuring Root Mean Square Error (RMSE), System Usability Scale (SUS) and NASA Task Load Index (TLX) metrics to assess accuracy usability, and cognitive workload, respectively.
}

\maketitle
\def\thefootnote{*}\footnotetext{The first author and second author are co-first authors.}\def\thefootnote{\arabic{footnote}}
\keywords{Ventriculostomy\and Low-Cost\and Augmented Reality \and Neurosurgery \and Mobile Computing\and Resource-limited Settings}

\section{Introduction}
Ventriculostomy, a common neurosurgical procedure, establishes a drainage pathway for cerebrospinal fluid (CSF) from the brain ventricles to a collection and monitoring system at bedside. This intervention aims to relieve excessive intracranial pressure within the skull caused by obstructed CSF flow.  The need for ventriculostomy arises when various pathological processes such as hemorrhages (e.g., aneurysms and vascular malformations), head trauma, tumours, spina bifida, hydrocephalus, or congenital issues block or impede normal CSF flow, CSF production, or its absorption. A recent global study found that traumatic brain injury (TBI) and the presence of hydrocephalus accounts for 45\% and 7\%, respectively, of all cases admitted for acute neurosurgical care.~\cite{dewan2018global}.

Placement of an external ventricular catheter (ventriculostomy) involves drilling a small hole in the skull and carefully inserting a thin, flexible catheter into the ventricle. Pre-operative CT scans along with well-known external cranial landmarks are typically used to guide placement for optimal accuracy. Relying solely on external landmarks in emergency situations can lead to misplacement in over 30\% of cases, and can potentially cause unwanted bleeding, inaccurate pressure readings, and ineffective drainage. These complications can also lead to longer hospital stays and increase in mortality~\cite{Edwards2015,Hagel2014}.

Image-guided neurosurgery (IGNS) improves any ventriculostomy procedure accuracy through use of pre-operative CT and MRI scans to provide real-time reference and spatial guidance, but its widespread use faces several challenges. The cost of acquiring these systems, ranging from USD 650,000 to over 900,000, can be a major barrier for many institutions~\cite{JSS4479}.Furthermore, effectively operating IGNS systems requires the expertise of specialized technicians skilled in planning and systems setup. Additionally, the bulky nature of the equipment, including the workstation and tracking camera, limits its use to large operating rooms, making it difficult to use these systems in other tighter settings like emergency rooms, wards, and ICUs (Intensive Care Units). These challenges limit IGNS use in low- and middle-income countries (LMICs) and remote communities. Additionally, a technician-intensive system is limited to elective and not emergency procedures.

Indeed, geographic disparities for timely and inexpensive EVD (External Ventricular Drain) placement exist. For example, Sub-Saharan Africa reports a much higher rate of infant hydrocephalus compared to other regions, with 750 new cases per 100,000 births compared to approximately 110 cases in Europe and the USA~\cite{GlobalhydrocephalusMichael}. This disparity in disease prevalence and access to advanced technology creates a concerning disparity in the quality of care between the resource-constrained settings with those high-resource settings. Our aim is to narrow this gap by developing a solution that is affordable, easily deployable, has a small footprint, and requires minimal expertise, making it suitable for use in LMICs, remote communities, and other resource-limited settings.

To address the challenges faced in translating surgical innovations into clinical practice ~\cite{reyes2022user}, we focus on designing a ventriculostomy guidance system through co-design with neurosurgeons and a focus on user experience. 
By applying user-centered design practices, we aim to create a system that aligns with both technological advances and the practical needs of surgeons, ensuring its usability and successful integration into daily surgical practice in resource-limited settings. Specifically, we present iSurgARy, a mobile system designed to operate exclusively on iPhones or iPads. iSurgARy leverages LiDAR for precise patient tracking, complemented by augmented reality (AR) for procedural guidance.

\section{Related Works}
Neurosurgical procedures are constantly evolving with the integration of innovative technologies such as robotics, mixed and virtual reality, 3D printing and intraoperative imaging methods. Hong \emph{et al.}~\cite{hong2023lowcost} introduced a mobile AR navigation system (MARNS) for precise transverse-sigmoid sinus junction location determination during retrosigmoid craniotomy. It demonstrates efficacy with a mean matching error of 2.88 mm (SD ± 0.69 mm), a positioning time of 279.71 seconds (SD ± 27.29 seconds), and was found to maintain bone flap integrity in 86.7\% of cases. In another study, de Almeida \emph{et al.}~\cite{DEALMEIDA2022e1261} addressed the cost and complexity of traditional neuronavigation systems. Their research presents a mobile-based AR solution for localizing points or landmarks on the scalp surface. In laboratory testing with a 3D phantom under optimal conditions, the system achieved an accuracy of 2.6 mm (SD ± 1.6 mm). Gorkem \emph{et al.}~\cite{GorkemYavas} investigated a 3D-printed marker-based AR system for intracranial tumor segmentation. It demonstrated high precision (0.5 to 3.5 mm targeting error), clinical feasibility, and cost-effectiveness, highlighting its potential for real-world application. In a separate study, L\'eger \emph{et al.} proposed MARIN, an iPad-based AR neuronavigation system~\cite{leger2020marin}, as well as NousNav~\cite{Leger2022}, a low-cost and an open-source neuronagivation system, as a solution for use in low-resource settings. NousNav uses low-cost off-the-shelf components, can be easily built for ~6k USD, features an intuitive interface and easy intraoperative control, and is robust and modular, making it a promising candidate for wider accessibility.

For ventriculostomy in particular, several mixed reality systems have been proposed to improve accuracy. Azimi et al.,~\cite{azimi2020interactive} developed an automated registration and trajectory planning system for ventriculostomy where a pointer equipped with a Vuforia marker was used for landmark-based registration. They achieved a 37\% improvement in accuracy for tip placement compared to manual registration methods, with a pointer tip-to-target distance of $10.96 mm$. In a similar study by Schneider \emph{et al.}~\cite{schneider2021augmented}, the authors used AR for ventriculostomy. They attached a Vuforia marker to the patient's head to guide the projection of 3D ventricle models onto the skull. A game controller was employed to align the holograms with the patient. Their system achieved a success rate of 68.2\% for ventriculostomy, with an average deviation of 5.2 mm from the planned trajectory. A subgroup showed significant improvement in results and precision after repeated attempts, suggesting a learning curve for using the AR system. The use of a rigid needle is assumed in AR-assisted medical procedures. Lin \emph{et al.}~\cite{lin2018holoneedle} explored the limitations of this approach. They developed a system utilizing the HoloLens to display segmented ventricles and the desired catheter trajectory, achieving a target registration error of $4.34 \pm 1.63$ mm and reducing catheter passes from $2.33 \pm 0.98$ to $1.07 \pm 0.258$ times. However, this method does not account for needle deflection, which can lead to inaccuracies. As highlighted in their study, incorporating real-time deflection data into AR systems could further improve precision in needle procedures. To improve freehand ventriculostomy accuracy, Ansari \emph{et al.}~\cite{NaghmehVentroAR} developed VentroAR, a HoloLens-based AR system. This system guides surgeons in locating ventricles, aiming to reduce the risk of complications associated with misplaced catheters. In a user study with 15 novices, VentroAR achieved an average gesture-based registration accuracy of $10.75 mm$ and a targeting accuracy of $10.64 mm$. While promising in terms of workflow and ease of use, the authors acknowledge the need for further accuracy improvements before clinical adoption. For a more comprehensive examination of mixed reality in ventriculostomy, readers are directed to the 2024 review by Alizadeh et al.~\cite{alizadeh2024virtual}.

Although tablet and smartphone systems have been developed for neurosurgery, to the best of our knowledge, they all either require external tracking systems, markers, or tags. Dogan et al.~\cite{Dogan2024} (2024) used a smartphone for neurosurgery but had issues with repeated manual “failed registrations”. \textcolor{reviewresponse}{Santos et al.~\cite{fernandes2021minimally}} developed a smartphone app to be used with a compass for neurosurgery procedures, yet their system required skin makers. De Almeida et al~\cite{DEALMEIDA2022e1261} developed a smartphone system, however their focus was on looking at the impact of lighting on registration. Furthermore, they used non-pro iPhones and thus were unable to take advantage of LiDAR for registration. MARIN~\cite{leger2020marin} uses an iPad yet requires the use of an external optical tracker. In comparison to these works, iSurgARy is novel as it requires the use of only a smartphone, uses LiDAR, and does not require additional trackers or markers on the patient. This simplifies the workflow in emergency and low-resource settings.

\section{Methodology}
iSurgARy was developed to be a small-footprint, low-cost image-guided neurosurgery (IGNS) system suitable for use in resource-limited and emergency settings. To achieve this, iSurgARy features a simple user interface (UI) designed to allow clinicians to easily select anatomical landmarks on the patient using the touch screen of mobile devices. These selected landmarks are then used in a rigid registration process to map the patient to their pre-operative scans.

\begin{figure*}[ht!]
    \centering
    \includegraphics[width=1.0\linewidth]{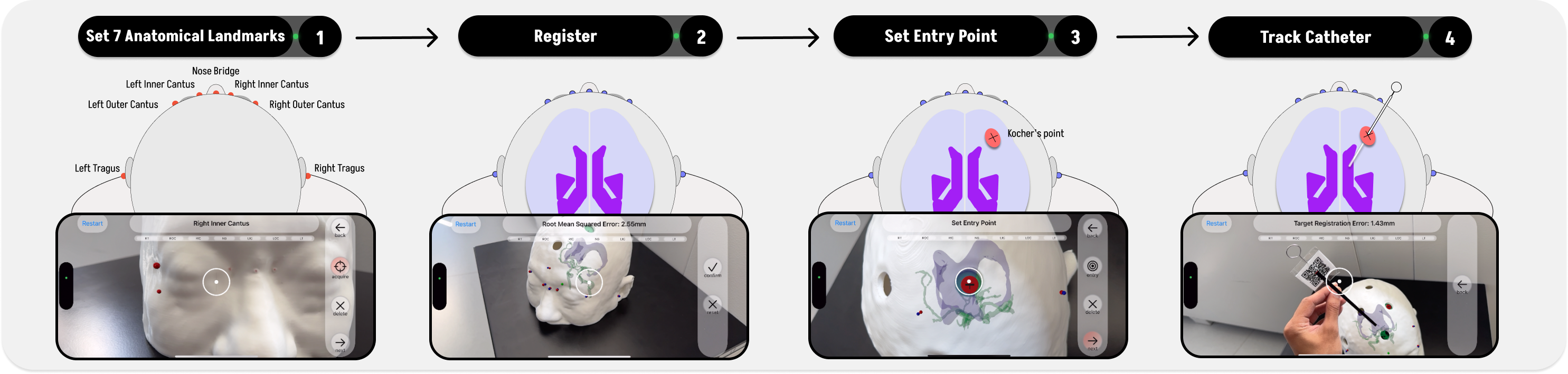}
    \caption{iSurgARy general workflow. (\emph{From left to right):} \textcolor{reviewresponse}{(1) The `acquire' button sets a reference landmark and displays a red sphere to visualize its placement in AR. Above, we display all landmarks from a head top point of view (2) The system displays the 3D model of the ventricles in AR with the registration's RMSE. The process of selecting landmarks more accurately can be repeated until the user is satisfied with the RMSE.  (3) The entry point placement feature enables the user to choose the catheter entry point on the surface of the head. We display the top view of Kocher's point, a common entry point for ventriculostomy. (4) The catheter tracking feature tracks a QR code image and renders the tip of the catheter in AR. The insertion of the catheter through the entry point from a head top point of view is shown.}}
    \label{fig:general work flow}
\end{figure*}

Once the registration process is complete, the clinician can visualize the ventricles (in the case of ventriculostomy) or other relevant 3D models (such as the cerebral vacsculature) in AR. This visualization aids in guiding the clinician to optimal EVD placement. Additionally, the catheter tracking tool enhances the clinician's spatial understanding of the distance between the catheter tip and the ventricles (see Figure \ref{fig:general work flow}).

\subsection{Development}
The current iSurgARy prototype utilizes an iPhone 15 Pro (MTUA3VC/A) running iOS 17.4 as the primary device. However, the application is also compatible with iPad Pro tablets that house the LiDAR sensor, offering greater flexibility in its use. Development is conducted within XCode Version 15.2 (15C500b) on a MacBook Air 15-inch, M2, 2023 with MacOS 14.3 (23D56), 16GB memory, and 1TB SSD. Additionally, a 3D-printed head phantom, designed for MRN system development and testing in neurosurgery, is used for alignment with corresponding digital ventricle and artery segments\cite{qi2024head}. iSurgARy was built for iOS using the Swift \cite{Swift} and C++\cite{cxxInterop} programming languages. It leverages Apple's ARKit\cite{ArKit}
and RealityKit\cite{RealityKit} frameworks to track and assign anatomical landmarks to 3D anchors and visualize the AR 3D model. The UI was implemented with SwiftUI\cite{SwiftUI}, while the asynchronous events are managed by the Combine framework\cite{Combine}. For unit conversion compatible with ARKit and transform matrix computation, Accelerate framework is utilized. On the C++ side, the iterative closest point algorithm (ItCP) and its helper functions rely on the Eigen library\cite{Eigen} and standard libraries like cmath, algorithm, and limits. 

\subsection{Workflow}

In the following section, we describe the user experience and workflow of using iSurgARy. Technical details about the tracking and registration implementation follow in the next sections. 

The application features a user-friendly interface designed for effortless registration and navigation. Users will load patient CT scans and select specific points on the scan for both landmark registration and surgical targeting. For instance, users can choose Koscher's point for burr hole localization and catheter targeting and may move the target entry point as deemed appropriate if there is cortical and ventricular distortion from blood or other space occupying lesions.

The app guides clinicians to place seven key anatomical landmarks on the patient, commonly used for patient-to-image registration in IGNS: the right tragus, the right outer canthus, the right inner canthus, the nose bridge, the left inner canthus, the left outer canthus, and the left tragus.

 During the registration process, the interface displays a text field at the top of the screen indicating the current landmark to target. A target cursor is centrally displayed on screen to aid in precise targeting. The right side of the interface contains an `acquire' button to confirm landmark placement relative to the center target cursor.  When the user presses the `acquire' button, a red spherical AR object is rendered on the device's camera feed at the aimed landmark. Once landmark placement is made, the user can either use the `delete' button to remove the AR landmark reference or choose a different anatomical target by using the `next' navigation buttons. The process of landmark placement for each target, guided by the text field at the top of the screen, is repeated until all seven key anatomical landmarks have been placed. The user can also navigate backwards and choose a previous anatomical target by using the `back' button. Technical details on landmark placement are discussed in Section \ref{AR}. Once all landmarks are placed, the `register' button aligns the landmarks to the AR model using an iterative closest point algorithm (ItCP), described in Section \ref{registration}. When the program completes registration, an AR view of the patient's internal structures for guidance is generated and the RMSE of the alignment is displayed at the top of the screen. If the user is dissatisfied with the alignment, they can easily re-select some or all of the anatomical landmarks for a more precise registration. If satisfied, the `confirm' button proceeds to the next phase.

In the EVD entry point placement phase, an entry point button enables the placement of a target for burr hole localization. If an entry point target is misplaced, a `delete' button allows the user to remove the saved entry point location and re-select an entry-point. After saving the entry point location and navigating to the next phase, using the `next' button, the target registration error (TRE) is displayed in the text field at the top of the screen (if the entry point was chosen on the pre-operative CT).

Catheter tracking is done using 2D image detection. Specifically, when the catheter tracking tool is within the camera frame, an AR view of a straight line is overlaid on the catheter from the QR code to the catheter tip, aiding in depth assessment during catheter insertion into the patient. Lastly, a `reset' button is available to remove all seven landmark targets placed by the user, providing a fast and convenient way to redo all landmarks, and reset the interface to the initial target stage where the user is asked to place the first landmark, the right tragus.

The interface allows clinicians to move around the patient from different angles, improving their 3D spatial awareness of internal structures by providing a better understanding of distance and direction between the catheter tip and the ventricle.

\begin{figure}[ht!]
    \centering
    \includegraphics[width=\linewidth]{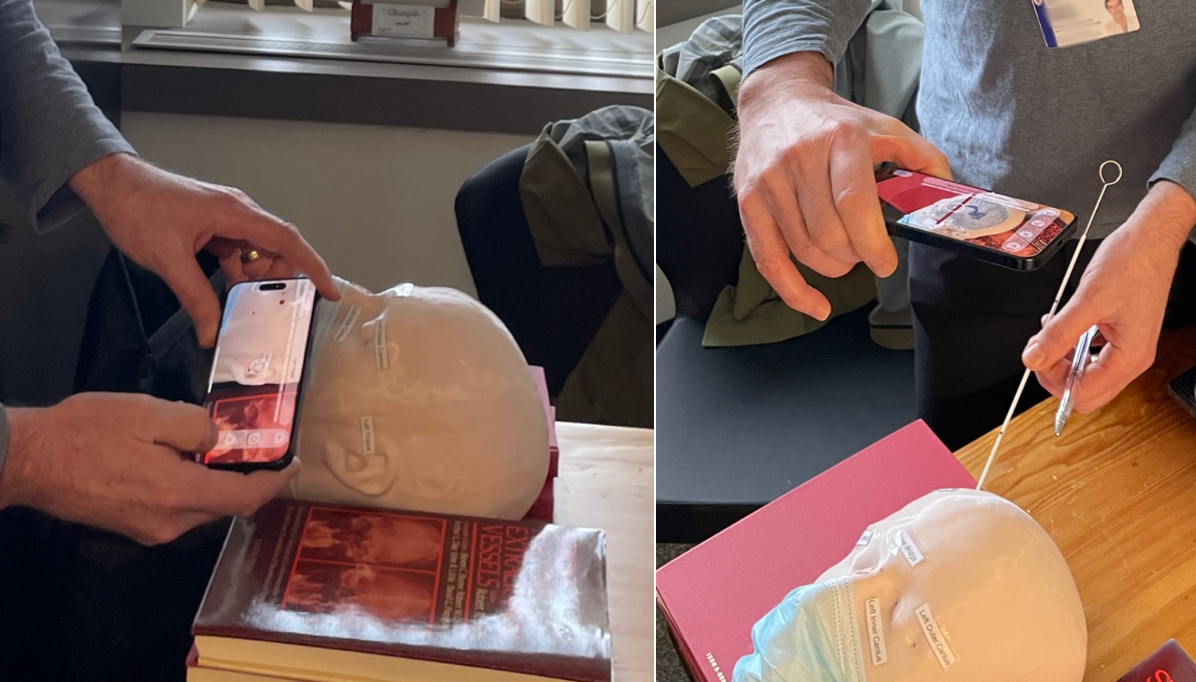}
    \caption{Left: Neurosurgeon performing registration by selecting the left tragus landmark. Right: Visualization of internal ventricular structure in AR and simulation of catheter insertion.}
    \label{fig:enter-label}
\end{figure}

\subsection{Patient-to-Image Registration}\label{registrtion}
For registration, i.e. mapping of patient space to image space, iterative closest point \textcolor{reviewresponse}{(ItCP)}~\cite{Zhang2014ICP} was used. ItCP aligns the landmarks placed in AR and the predefined landmarks from the patient's CT scan. Our ItCP implementation processes coordinates of both AR-placed landmarks and predefined landmarks, producing a scale factor, a rotation matrix, and the RMSE. 

To visualize the aligned internal structures in AR, the 3D model is anchored at the centroid of the AR landmarks, with its origin as the mean of the predefined landmarks. The scale factor derived from the ItCP process is applied to the loaded 3D model. The rotation matrix is then used to adjust the orientation of the 3D model ensuring optimal alignment of the landmarks.

\subsection{Augmented Reality Visualization}\label{AR}
Traditional IGNS systems present navigation information on a computer outside the sterile operating region. This requires the surgeon to shift their gaze away from the patient and surgical field for guidance information and mentally map it to the patient's anatomy. This process can be time-consuming, cognitively demanding, and prone to error~\cite{kersten-oertel2013state, asadi2024decade}. Furthermore, it results in a disconnect between where the surgeon is working and where they are looking for guidance. This constant shifting of attention can negatively influence the successful completion of a surgical task~\cite{leger2017quantifying}. 

To address these issues, \textcolor{reviewresponse}{we use AR} to superimpose the patient’s ventricular anatomy onto the head and track the tip of the catheter. In iSurgARy, ARKit provides world tracking capabilities, utilizing the devices' camera and motion sensors to monitor its position and orientation in the real world\cite{ARKitWorldTracking}. \textcolor{reviewresponse}{In an ARKit ARSession\cite{ARSession}, feature points are detected within the camera's image space and used to represent key details of the real-world environment. These feature points are then compiled into an ARPointCloud, which is a collection of 3D coordinates in the world space that ARKit uses for various tracking tasks. This point cloud is crucial for continuously updating fixed anchor points in the environment that serve as reference frames for placing and tracking virtual objects in real-time as the camera moves. ARKit uses this data to ensure that virtual content remains accurately positioned and aligned with the physical world across each frame. Additionally, the housed LiDAR scanner enhances depth tracking and anchor placement precision, we access the depth data using ARDepthData\cite{ARDepthData}. When the user presses the `acquire' button for placing a landmark, we use ARCamera\cite{ARCamera} to capture the position and orientation of the camera, and we use ARDepthData to determine the distance between the device and the subject being targeted. From this, we compute the spatial coordinates of the landmarks from each capture and pass it into the ItCP algorithm in order to align our 3D model to the real-world environment. RealityKit is employed to render the 3D content, such as the virtual 3D spheres for the placed landmarks and the 3D model of the ventricles. To track the tip of the catheter, we pass a QR code image as a reference image to the detectionImages property of the world-tracking configuration\cite{imageDetection}. When the ARsession recognizes the QR code reference image, an ARImageAnchor\cite{ARImageAnchor} is added to the detected image. Once detected, a rectangular AR object is rendered from the detected image extending to the tip of the catheter as shown in rightmost image of Figure ~\ref{fig:general work flow}.}

\section{User Study}
To evaluate the iSurgARy app in a laboratory setting, we conducted a two-phase user study. The first phase involved a preliminary study with novice participants, focusing on assessing the usability and learnability of the app. Feedback from this phase was used to refine the application. In the second phase, we engaged clinical and HCI experts to evaluate usability and  performance. This phase included measuring RMSE to assess accuracy, SUS for system usability and the NASA TLX to evaluate cognitive workload. 

\subsection{Novice User Study} 
For the novice study, we had 10 university participants (5 female and 5 male) who were graduate students in engineering and science fields. They performed landmark registration and patient tracking and completed the SUS\cite{SUS} to provide feedback on the app’s usability. We started with a brief introduction of the app to each participant. We explained that the accuracy of task completion was not the primary concern. Instead, the focus was on how easily the participants could follow the app instructions and navigate the user interface to complete the tasks. This approach ensured that we received meaningful feedback on the usability and learnability of the app, allowing us to identify and address any issues.

\begin{figure}
        \centering
        \includegraphics[width=\linewidth]{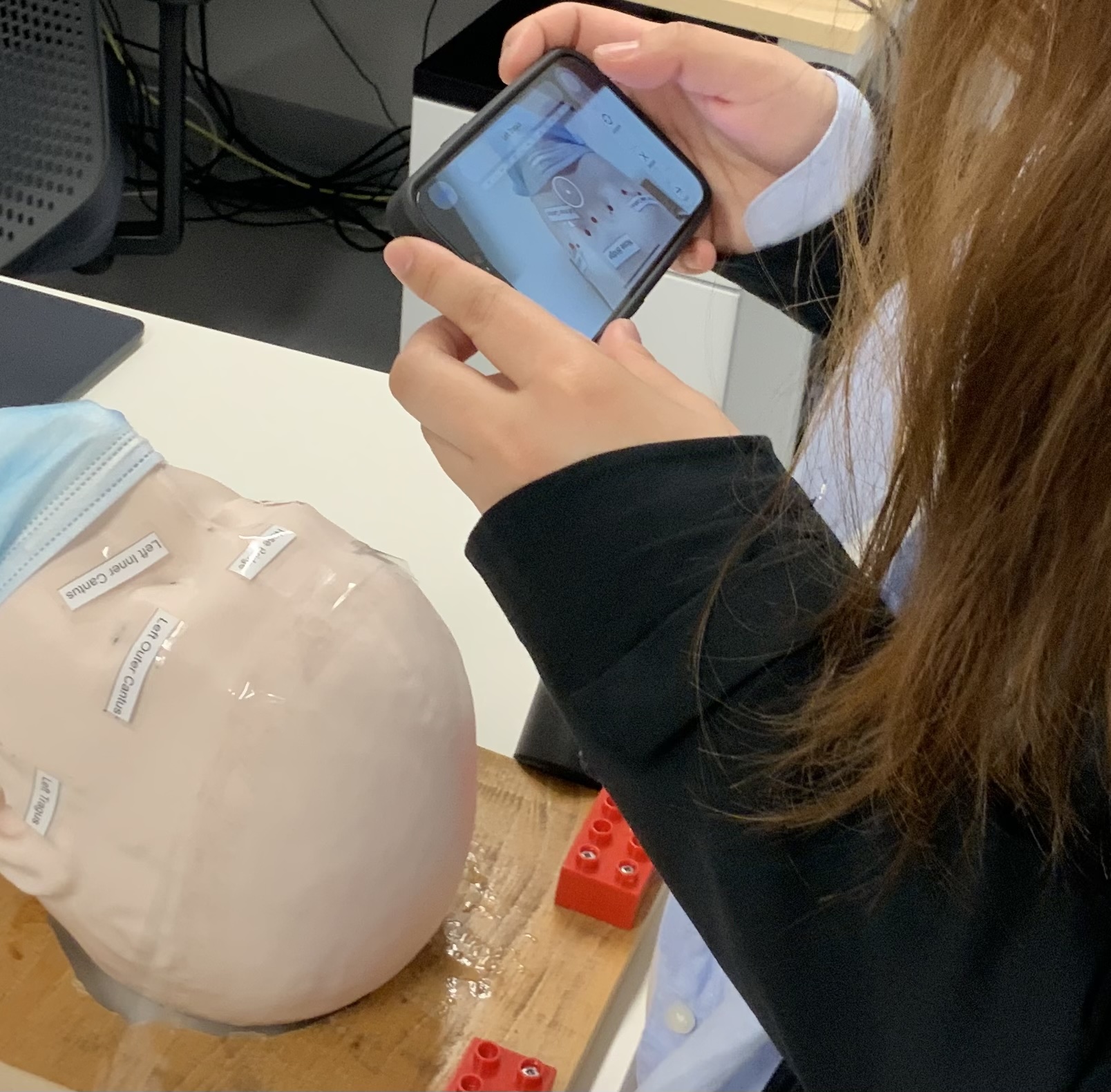}
        \caption{Participant uses the aim cursor positioned at the center of the screen to select landmarks.}
        \label{fig:experiment}
    \end{figure}

\subsection{HCI/Clinical Domain Expert Study} 
In the second phase of our evaluation, we focused on assessing both the usability and learnability of iSurgARy, as well as its accuracy and cognitive load when used by domain knowledge users. This study involved 11 participants, including nurses, clinicians, and HCI experts. The demographics and backgrounds of these domain experts are presented in Table~\ref{tab:Respondents_demographics}.
We had a gender split of almost 50\%, with the following backgrounds: 
\begin{itemize}
    \item A neurosurgeon with over 25 years of experience
    \item A clinician with a Master’s degree focused on IGNS
    \item Three nurses working at a neurological institute, including two with OR/ER experience and one nurse manager
    \item Three HCI researchers
    \item One HCI and mobile app developer
    \item One IGNS researcher/developer
\end{itemize}

All participants had relevant or related experience, ensuring the feedback we received was well-informed and applicable to real-world scenarios. The study concluded by collecting qualitative feedback on the app's usability. This feedback guided further improvements in the application's design.

\begin{table}[h]
\caption{Expert Participants' Demographics and Background}
\label{tab:Respondents_demographics}
\centering
\begin{tabular}{p{0.48\linewidth} >{\raggedleft\arraybackslash}p{0.2\linewidth} >{\raggedleft\arraybackslash}p{0.2\linewidth}}
\toprule
\textbf{Demographics} & \textbf{Number} & \textbf{Percentage} \\
\midrule
\textbf{Age (years)} & {n = 11} & \\
\quad 18 - 24 & 1 & 9.1\% \\
\quad 25 - 34 & 5 & 45.5\% \\
\quad 35 - 44 & 4 & 36.4\% \\
\quad 65+ & 1 & 9.1\% \\
\midrule
\textbf{Gender} & {n = 11} & \\
\quad Male & 6 & 54.5\% \\
\quad Female & 5 & 45.5\% \\
\midrule
\textbf{Expertise} & {n = 11} & \\
\quad HCI Experts & 5 & 45.5\% \\
\quad Clinical Domain Experts & 6 & 54.5\% \\
\midrule
\textbf{Education} & {n = 11} & \\
\quad Bachelor's Degree & 2 & 18.2\% \\
\quad Professional Degree & 2 & 18.2\% \\
\quad Master's Degree & 4 & 36.4\% \\
\quad Doctorate & 3 & 27.3\% \\
\midrule
\textbf{Years in Field} & {n = 11} & \\
\quad 1 year & 2 & 18.2\% \\
\quad 2 - 6 years & 4 & 36.4\% \\
\quad 6 - 15 years & 4 & 36.4\% \\
\quad 30+ years & 1 & 9.1\% \\
\bottomrule
\hline
\hfill
\end{tabular}
\label{tab:DemoResults}
\end{table}

\section{Results}

\subsection{Novice User Study} For system usability, we used the SUS, a questionnaire employed to evaluate the usability of products and services. The results of the SUS are presented in Table~\ref{tab:SUS_table_novice}. The average SUS was 81.52\% indicating an easy-to-use system (Note: The average SUS score is 68\% and scores up to 70\% are considered good). Furthermore, comments from the participants indicated that they felt the system was easy to learn, understand and use. 

\begin{table*}[h!]
    \caption{Novice User Study: The 10 questions of the System Usability Scale (SUS) with the average score from 1 (strongly disagree) to 5 (strongly agree). \textcolor{reviewresponse}{Rows in green correspond to positive statements, whereas rows in white correspond to negative statements.}}
    \centering
    \setlength{\tabcolsep}{15pt} 
    \renewcommand{\arraystretch}{1.3} 
    \begin{tabular}{p{0.8\textwidth} c}
        \toprule
        \textbf{System Usability Scale for Novice User Study} & \textbf{Average} \\
        \midrule
        \rowcolor{mediumgreen}I think that I would like to use this system frequently. & 4.3 \\
        I found this system unnecessarily complex. & 1.4 \\
        \rowcolor{mediumgreen} I thought this system was easy to use. & 4.4 \\
        I think I would need help to use this system. & 2.5 \\
        \rowcolor{mediumgreen} I found the various functions in this system were well integrated. & 4.2 \\
        I thought there was too much inconsistency in this system. & 2.1 \\
        \rowcolor{mediumgreen} I would imagine that most people would learn to use this system very quickly. & 4.7 \\
        I found this system very cumbersome/awkward to use. & 1.36 \\
        \rowcolor{mediumgreen} I felt very confident using this system. & 3.82 \\
        I needed to learn a lot of things before I could get going with this system. & 1.45 \\
        \midrule
        \textbf{Mean $\pm$ SD} & \textbf{\textcolor{reviewresponse}{81.52 $\pm$ 0.4}} \\
        \bottomrule
    \end{tabular}
    \label{tab:SUS_table_novice}
\end{table*}

\subsection{HCI/Clinical Domain Expert Study} 
For evaluating system usability in our second study, we used SUS, the results of which are summarized in Table~\ref{tab:SUS_table_expert}. The average SUS was 80.95\%, indicating that the system was perceived as easy to use by our expert domain knowledge participants. 

\begin{table*}[h!]
    \caption{Expert User Study:The 10 questions of the System Usability Scale (SUS) with the average score from 1 (strongly disagree) to 5 (strongly agree). \textcolor{reviewresponse}{Rows in green correspond to positive statements, whereas rows in white correspond to negative statements.}}
    \centering
    \setlength{\tabcolsep}{15pt} 
    \renewcommand{\arraystretch}{1.3} 
    \begin{tabular}{p{0.8\textwidth} c}
        \toprule
        \textbf{System Usability Scale for Expert User Study} & \textbf{Average} \\
        \midrule
        \rowcolor{mediumgreen} I think that I would like to use this system frequently. & 4.36 \\
        I found this system unnecessarily complex. & 1.36 \\
        \rowcolor{mediumgreen} I thought this system was easy to use. & 4.27 \\
        I think I would need help to use this system. & 1.82 \\
        \rowcolor{mediumgreen} I found the various functions in this system were well integrated. & 4.55 \\
        I thought there was too much inconsistency in this system. & 1.55 \\
        \rowcolor{mediumgreen} I would imagine that most people would learn to use this system very quickly. & 4.73 \\
        I found this system very cumbersome/awkward to use. & 2.1 \\
        \rowcolor{mediumgreen} I felt very confident using this system. & 3.5 \\
        I needed to learn a lot of things before I could get going with this system. & 2.2 \\
        \midrule
        \textbf{Mean $\pm$ SD} & \textbf{\textcolor{reviewresponse}{80.95 $\pm$ 0.4}} \\
        \bottomrule
    \end{tabular}
    \label{tab:SUS_table_expert}
\end{table*}

\begin{figure}[h!]
    \centering
    \includegraphics[width=\linewidth]{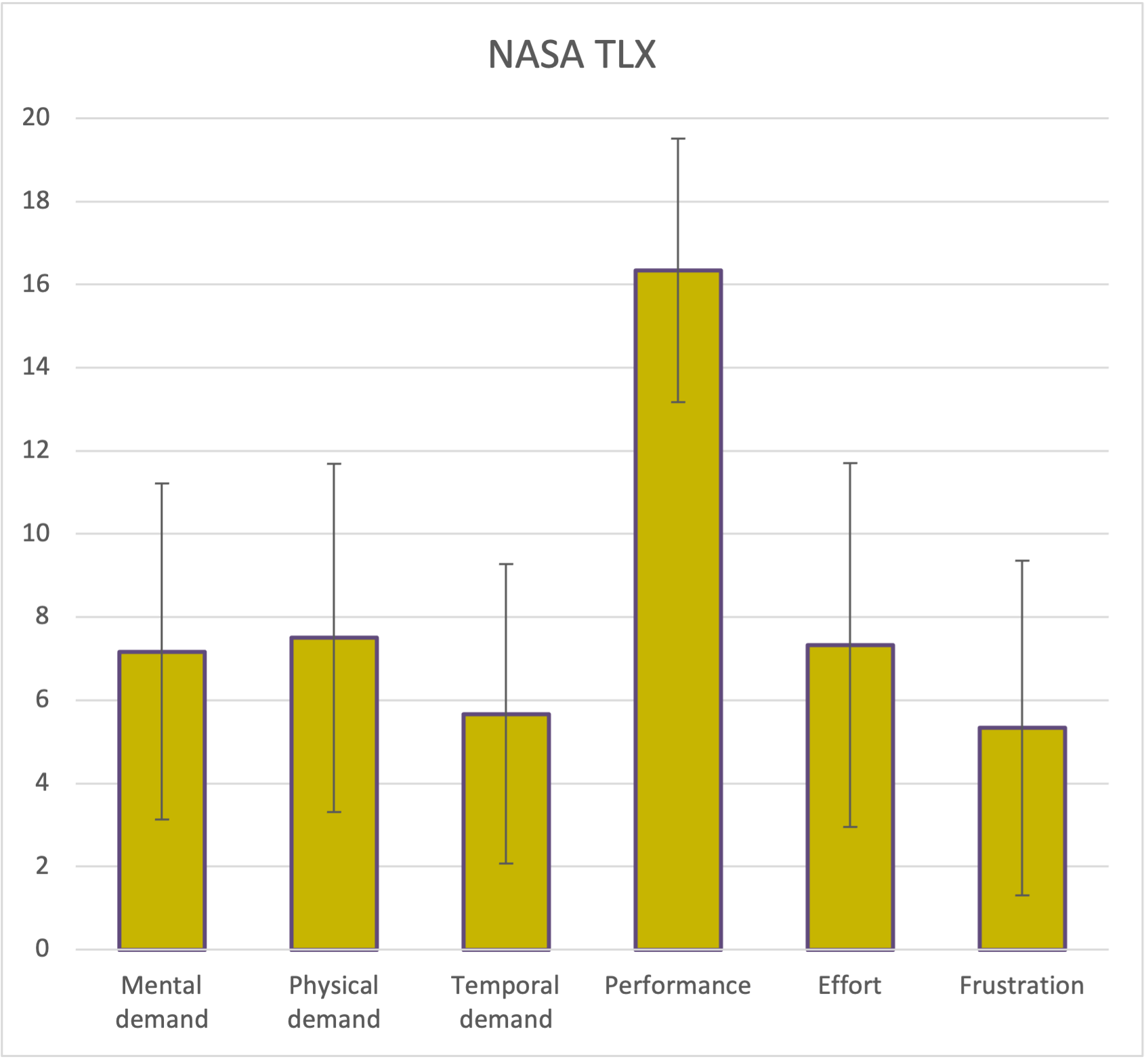}
    \caption{NASA TLX Scores for each dimension of workload.}
    \label{fig:nasa_tlx_scores}
\end{figure}

The NASA Task Load Index (NASA TLX) was used to assess the subjective workload experienced by participants. The mean scores (with standard deviations) for the dimensions were as follows: Mental Demand, 7.17 (±4.04); Physical Demand, 7.5 (±4.19); Temporal Demand, 5.67 (±3.60); Performance, 16.33 (±3.17); Effort, 7.33 (±4.38); and Frustration, 5.33 (±4.03) (see  Figure~\ref{fig:nasa_tlx_scores}). Among these dimensions, performance had the highest mean score, suggesting that participants felt they had to work hard to get a good level of performance, in this case a low RMSE. The overall workload, based on the NASA TLX scores is 41.1 on a 100-point scale suggesting a moderate workload. The variability in standard deviations across the dimensions reflects differing levels of consensus among participants regarding their experience. 

In terms of accuracy, on average, we found an RMSE of $2.54 mm \pm 0.46$. These findings align with previous works~\cite{DEALMEIDA2022e1261,hong2023lowcost}. Furthermore, according to our experts as the ventricles are a large target, an RMSE of < 5 mm is generally acceptable. 

\subsubsection{Qualitative Feedback}
To ensure the application's scope aligned with the needs of neurosurgeons, we collaborated closely with them during the development process. One of the neurosurgeons who participated in our expert user study provided particularly encouraging feedback. Comments included: \emph{"Wow. I think you have the gist of it. Simple and quick registration and very responsive tracking. Sign me up. When would you like to test this in a clinical setting?"}, \emph{
"I would like to use the iPad, but I think the residents will prefer the iPhone form factor as they can just pull it out of their pocket."} and \emph{"Superb user-friendly and easy-to-use tool for the task at hand. It will be a valuable adjunct towards the safe and timely placement of EVDs."}

Other comments included: \emph{"Another anchor point that isn't co-planar with rigid points to better account for out of plane rotations"} and \emph{"Try using multiple selections of the same point to both assess intra-user variability in point selection and to use centroid of chosen points as anchor point for registration"}, suggesting that improvements could be made on the registration procedure.

\begin{figure}
    \centering
    \includegraphics[width=\linewidth]{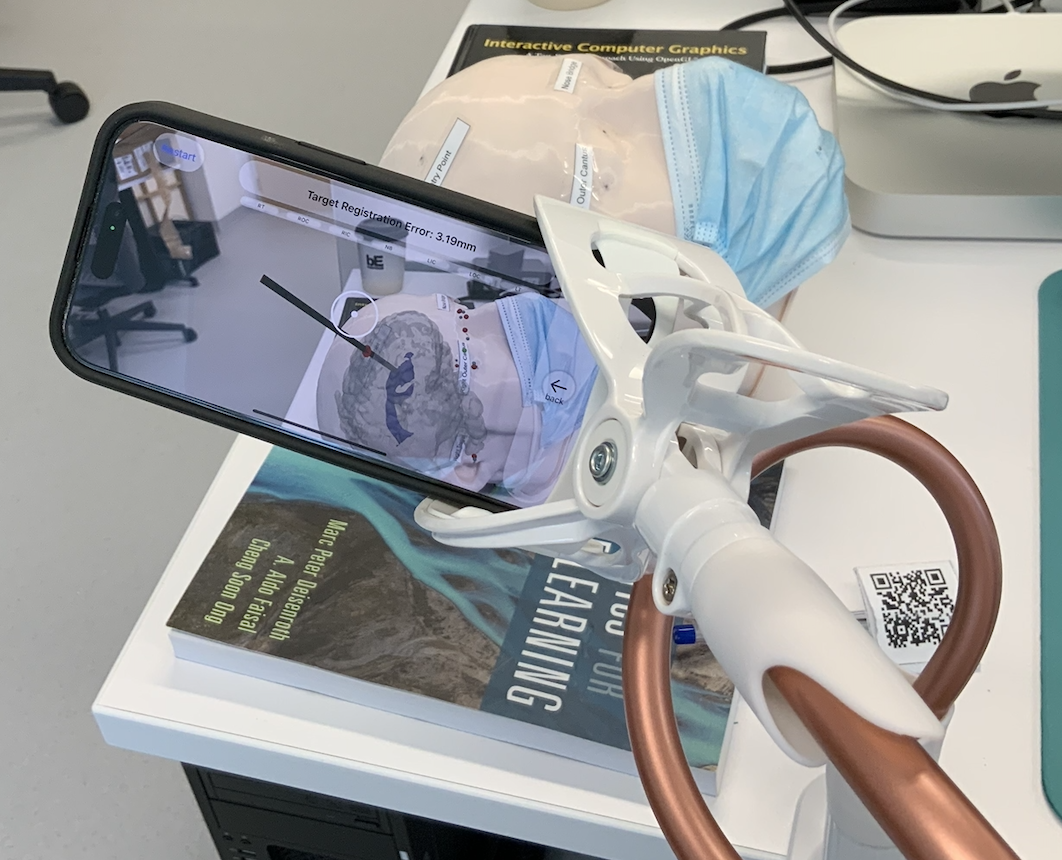}
    \caption{After feedback from clinicians we began using a mobile phone holder to improve ergonomic issues.}
    \label{fig:ergonomic phone holder}
\end{figure}

Based on the various feedback, we have determined an ergonomic solution that would work in resource-limited emergency and ICU settings. First, an iPhone/iPad holder that would be clamped to the bed during the procedure would allow the user to have both hands available for \textcolor{reviewresponse}{catheter} placement (see Figure~\ref{fig:ergonomic phone holder}). Next, simple solutions for stabalizing the patient's head were discussed. \textcolor{reviewresponse}{Lastly, in future work,} to ensure that the \textcolor{reviewresponse}{catheter} can be tracked without the need for the QR code, we will focus on developing computer vision techniques. 

\section{Discussion}
iSurgARy prioritizes affordability, aiming to make advanced navigation technology accessible to a wider range of medical institutions while maintaining accuracy. Unlike traditional IGNS systems, which can cost thousands of dollars due to sophisticated tracking cameras and dedicated computers, iSurgARy utilizes commercially available and affordable mobile devices. While there is an initial investment associated with these devices, it represents a significantly more economical option. The cost-effectiveness, combined with the portability and ease of use of our system makes it a much better navigation option for resource-constrained settings. 

iSurgARy combines both human expertise and machine capabilities, and in doing so, this introduces potential sources of error for which the user should be aware. The initial placement of virtual landmarks is a key step for accurate guidance, but it relies on the user's judgment and precision. This can lead to inaccuracies that may significantly affect subsequent stages, particularly alignment. Errors can also arise from the ItCP algorithm and the ARKit framework. The ItCP algorithm, while robust in aligning 3D structures, can struggle with sparse data, uneven distributions of landmarks, or large initial misalignment that exceed its convergence criteria. The ARKit framework, used for real-world environment tracking, has its own limitations due to sensor accuracy and environmental factors. The iPhone/iPad uses an accelerometer and a gyroscope, and their precision limitations can manifest as inaccuracies in landmark tracking, potentially causing misalignment between the real world and the augmented overlay. In future work, we will consider capturing a point cloud of the patient with LiDAR and performing a more dense point-based registration. 

Despite a modest user study sample size, we believe our findings provide reliable results about the usability and applicability of the system. Our second study aimed at using domain experts (mobile app developers, HCI experts and clinicians) to determine the usability of the system as human-computer interaction (HCI) research suggests that a sample of even just 3-5 system evaluators can identify ~75\% of usability issues. Furthermore, the inclusion of double experts (in this case a clinician who was also an IGNS researcher) may identify more issues~\cite{abulfaraj2020detailed, cho2022assessing}. Feedback from the neurosurgeon, clinician with IGNS research background and an IGNS researcher and developer, as well as our HCI experts not only validated our approach but also helped us identify areas for improvement in the next iteration of our application. This collaboration underscores the importance of iterative development and expert input in refining and enhancing the usability and functionality of medical applications.

The next step of this work will be to further test iSurgARy with neurosurgeons and residents. This testing will be important in evaluating the robustness and accuracy of our system in real-world surgical settings, determining its practical viability and clinical effectiveness in achieving precise EVD placement. Furthermore, we plan to compare our smartphone/tablet system accuracy, usability, and ergonomics with head-mounted displays (HMDs) like the Apple Vision Pro or Microsoft HoloLens. These devices potentially offer a hands-free experience and potentially better immersion, which could be advantageous for complex procedures. However, weight, battery life, and maintaining sterility remain challenges. Our future research will explore these trade-offs to identify the most effective AR solution for surgical applications.

\section{Conclusion}
In this work, we aimed to develop a practical system with an intuitive interface that could be used in an ICU, emergency room and/or resource-limited settings. By working closely with neurosurgeons in defining the app’s scope that aligns with their needs, iSurgARy addresses a critical need for a low-cost, portable, and user-friendly IGNS system in resource-limited settings and emergency situations. Ventriculostomy, a common neurosurgical procedure, is particularly suited to such a system. Our initial prototype demonstrates promising accuracy and sets the stage for continued refinement and clinical evaluation. Successful implementation of iSurgARy has the potential to significantly improve the accuracy of EVD placement, leading to reduced healthcare costs, mortality and morbidity rates.

\fundingandinterests{The work described in this paper was funded by XXX
Conflict of interest: None declared.}

\bibliographystyle{acm}
\bibliography{references}

\end{document}